\documentclass[twocolumn,showpacs,preprintnumbers,amsmath,amssymb]{revtex4}
\usepackage{graphicx}% Include figure files
\usepackage{dcolumn}% Align table columns on decimal point
\usepackage{bm}% bold math
\begin{document}

\preprint{APS/123-QED}

\title{Lagrangian dispersion and heat transport in convective turbulence}
% Force line breaks with \\

\author{J\"org Schumacher$^{1,2}$}
\affiliation{$^1$Department of Mechanical Engineering, Technische Universit\"at Ilmenau, D-98684 Ilmenau, Germany,\\
             $^2$International Collaboration for Turbulence Research (ICTR)}

\date{\today}% It is always \today, today,

\begin{abstract}
Lagrangian studies of the local temperature mixing and heat transport in turbulent Rayleigh-B\'{e}nard convection are presented, 
based on three-dimensional direct numerical simulations. Contrary to vertical pair distances, the temporal growth 
of lateral pair distances agrees with the Richardson law, but yields a smaller
Richardson constant due to correlated pair motion in plumes. Our results thus imply that Richardson dispersion is also found in 
anisotropic turbulence. We find that extremely large vertical accelerations appear less frequently than lateral ones and 
are not connected with rising or falling thermal plumes. The height-dependent joint Lagrangian statistics of vertical 
acceleration and local heat transfer allows us to identify a zone which is dominated by thermal plume mixing.  
\end{abstract}

\maketitle

{\em Introduction.} The interplay of turbulent temperature and velocity fields in thermal convection is present in many systems
reaching from astro- and geophysical flows \cite{Cattaneo2003,Stevens2005} to chemical engineering and
indoor ventilation \cite{Linden1999}. Almost all experimental and numerical studies on thermal convection have been
carried out in the Eulerian frame of reference where the velocity and the temperature are analysed at fixed spatial positions and
the mean turbulent heat transfer is measured with increasing precision for ever larger Rayleigh numbers \cite{Castaing1989,Niemela2000,Ahlers2005}. Not much is however known about the local mechanisms of the transport of heat and momentum.  
Heat transport fluctuates locallly since it is arranged in coherent thermal plumes. They detach permanently from the thermal boundary layers, 
carry blobs of hotter or colder fluid into the bulk and cause an efficient temperature mixing. The complementary Lagrangian 
description in which turbulent fields are monitored along the tracks of fluid parcels 
\cite{Mann2000,LaPorta2001,Mordant2001,Yeung2002,Boffetta2002,Biferale2004,Bourgoin2006,Grauer2007} could thus provide 
exactly the missing 
insights on the local mixing and transport mechanisms in convective turbulence. For example, the frequency and seasonal 
variability of thermal plumes affects the transport and concentration of phytoplankton in the upper ocean and has an impact
on the carbon cycle \cite{Kortzinger2004,Riser2008}. Vertically transported graupel particles or water droplets 
in deep-convective cumulus clouds determine how much precipitation falls out \cite{Blyth1993,Shaw2003}.

In this Letter, we study the Lagrangian mixing of turbulent Rayleigh-B\'{e}nard convection in
three-dimensional direct numerical simulations. Two aspects are in the focus of the present investigation. Firstly, we provide a 
detailed analysis of the acceleration and dispersion properties of Lagrangian tracers in convection. Caused by the interplay of 
buoyancy and gravity, Rayleigh-B\'{e}nard turbulence is inhomogeneous and anisotropic. Consequently, significant 
differences between the vertical and lateral transport properties are found. Secondly, we will take a "Lagrangian fingerprint'' of thermal 
plumes. We therefore explore the statistics of local heat transfer and its relation to the local acceleration along
the Lagrangian tracer tracks, all this as a function of height. This allows us to identify the regions in the convection
cell which are dominated by 
the plumes and in which consequently enhanced vertical temperature mixing is present \cite{Castaing1989}.
Our studies extend recent Eulerian investigations on the plume structure \cite{Shishkina2007,Zhou2007} and an 
experiment with neutrally buoyant smart
temperature sensors which was able to probe Lagrangian bulk properties in high-Rayleigh-number convection on scales down to 17
times the thermal boundary layer thickness \cite{Gasteuil2007}.          

%-------------------------------------------------------------------------------
\begin{figure}
\includegraphics[width=8cm]{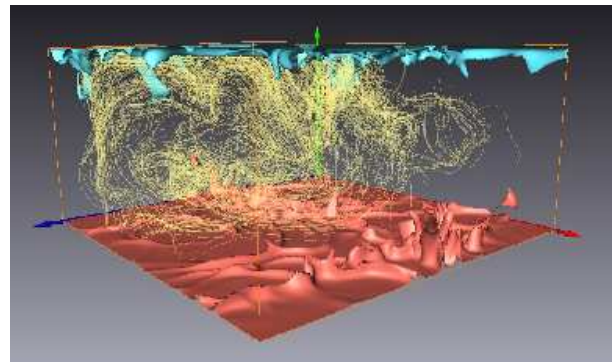}
\caption{(Color online) Instantaneous snapshot of two isosurfaces of the total temperature $T$ (top=cold, bottom=hot).
Thermal plumes which detach from both planes can be observed. 
Streamlines of the velocity field are also shown.} 
\label{fig1}
\end{figure}
%-------------------------------------------------------------------------------
\begin{figure}
\includegraphics[width=7cm]{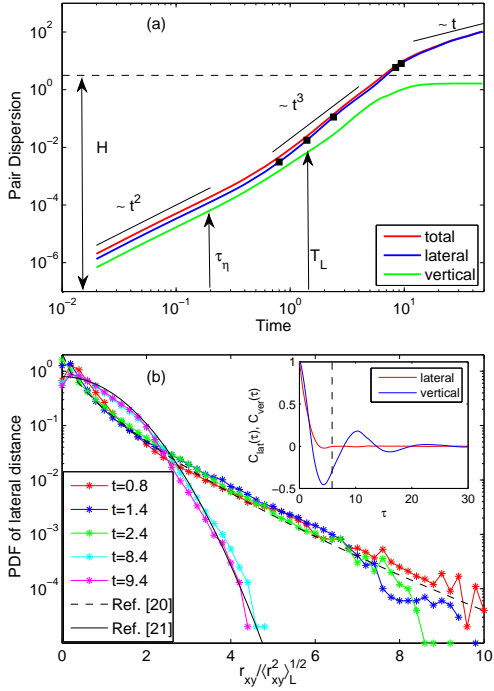}
\caption{(Color online) Particle pair dispersion in turbulent convection. (a): Pair dispersion 
$\langle[\bm{\Delta}(t)-\bm{\Delta}(0)]^2\rangle_L$ as a function of time $t$. The vector ${\bm\Delta}$ is the 
distance vector between both tracer positions within a pair. The vertical, lateral
and total dispersion are compared. The height of the convection cell, the Kolmogorov time, $\tau_{\eta}$, the 
lateral Lagrangian integral time, $T_L$, are also indicated.
(b): Probability density function (PDF) of the lateral dispersion taken at four times, see symbols in (a). 
For comparison, we include the Gaussian prediction by Batchelor \cite{Batchelor1950} and the Richardson
prediction \cite{Richardson1926} (with different fit factors). The inset shows the Lagrangian
velocity autocorrelation functions for the lateral, $C_{lat}(\tau)$ and vertical , $C_{ver}(\tau)$ components.
The vertical dashed line corresponds with the travel time $H/u_{z,rms}$.} 
\label{fig2}
\end{figure}
%-------------------------------------------------------------------------------

{\em Numerical simulations.}
The Boussinesq equations for an incompressible flow and the advection-diffusion 
equation for the temperature field, are solved by a standard pseudospectral method for 
the three-dimensional case. The equations are given by
\begin{eqnarray}
{\bm\nabla\cdot \bm u}&=&0\,,\\
\frac{\partial {\bm u}}{\partial t}+({\bm u\cdot\bm\nabla}){\bm u}&=&
-{\bm\nabla} p +\nu {\bm\nabla}^2{\bm u}+\alpha g \theta {\bm e}_z\,,\\
\frac{\partial \theta}{\partial t}+({\bm u\cdot\bm\nabla})\theta &=&
\kappa {\bm\nabla}^2\theta+u_z \frac{\Delta T}{H}\,.
\end{eqnarray}
Here, ${\bm u}$ is the turbulent velocity field, $p$ the pressure field and $\theta$ the temperature field. 
The system parameters are:  gravity acceleration $g$, kinematic viscosity $\nu$, thermal diffusivity $\kappa$, 
vertical temperature gradient $\Delta T/H$, and thermal expansion coefficient $\alpha$. The total temperature 
field $T$ is decomposed into a linear background profile and a fluctuating field $\theta$
%----------------------------------------------------------------------------------
\begin{equation}
T({\bm x},t)=-\frac{\Delta T}{H}(z-H/2)+\theta({\bm x},t)\,.
\label{reynolds0}
\end{equation}
%----------------------------------------------------------------------------------
Since $T=\pm \Delta T/2$ at boundaries $z=0$ and $z=H(=\pi)$, the boundary condition $\theta=0$ follows. 
Here, $\Delta T >0$. The dimensionless control parameters are the Prandtl number $Pr=\nu/\kappa=0.7$, 
the Rayleigh number $Ra=\alpha g H^3 \Delta T/(\nu\kappa)=1.2 \times 10^8$, and the aspect ratio 
$\Gamma=L/H=2$ with $L=L_x=L_y=2\pi$. Figure 1 illustrates a typical snapshot of the flow and temperature fields.
In the vertical direction $z$, free-slip boundary conditions are used.  Laterally in $x$ and $y$, 
periodic boundary conditions are applied. The simulation box is resolved by $512\times 512\times 257$
grid points which results in a spectral resolution of $k_{max}\eta_K=2.2$. The Taylor microscale
The Reynolds number is $R_{\lambda}=135$ and the Nusselt number $Nu=58.05\pm 0.98$. Time-stepping is done 
by a second-order predictor-corrector scheme. We track $N=10^6$ Lagrangian tracers simultaneously
with the flow. Initially, they are arranged in pairs with a separation of 1.5 grid cell widths. Intergrid
velocities are obtained by trilinear interpolation.  
%-------------------------------------------------------------------------------
\begin{figure}
\includegraphics[width=8cm]{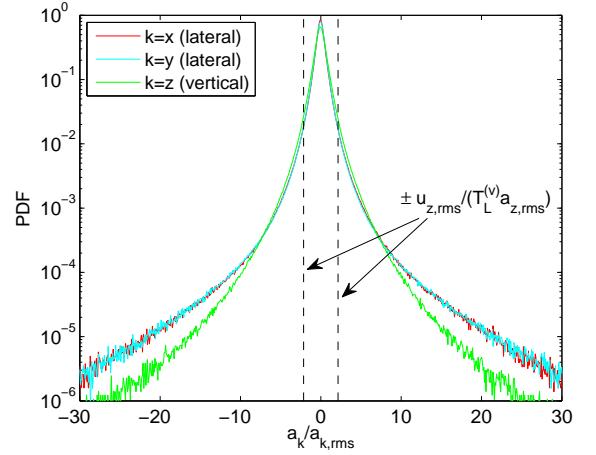}
\caption{(Color online) Probability density function (PDF) of the acceleration vector components. Components
are normalized by their corresponding root-mean-square values (rms). The skewness of all three distributions
vaires between -0.2 and -0.09. The dashed vertical lines mark the characteristic acceleration $u_{z,rms}/T_L^{(v)}$.} 
\label{fig3}
\end{figure}
%-------------------------------------------------------------------------------

{\em Particle dispersion.} Figure 2a reports our findings on the pair dispersion which is given by 
$R^2(t)=\langle [{\bm\Delta}(t)-{\bm\Delta}(0)]^2\rangle_L$. The statistics is taken over $N/2$ tracer 
pairs and denoted as $\langle\cdot\rangle_L$. Each pair is described by the particle distance vector
${\bm\Delta}={\bm x}_2-{\bm x}_1$.  Interestingly, after the initial ballistic growth with $t^2$ 
a crossover to a Richardson-like dispersion with $R^2(t)=C_2\langle\epsilon\rangle t^3$  is observed, similar to 
isotropic fluid turbulence \cite{Mann2000,Boffetta2002,Richardson1926}. Here, $\langle\epsilon\rangle$ 
is the mean energy dissipation rate. In contrast to the isotropic case, the Richardson constant is 
however reduced by two third and is given by $C_2=0.16$ in the present study \cite{Mann2000,Boffetta2002}. Recall, that the 
turbulence is inhomogeneous with respect to the vertical direction. Therefore, we decompose the distance vector into 
a lateral and vertical part, ${\bm\Delta}={\bm r}_{xy}+r_z{\bm e}_z$. While the vertical distances cannot 
grow beyond the cell height $H$, the lateral one takes over and makes up the whole large-time growth of $R^2(t)$ including
the Richardson scaling.
Eventually, the lateral pair motion becomes decorrelated and changes to the diffusive limit with $R^2(t)\sim t$
(see Fig. 2a). 

In Fig. 2a, the viscous Kolmogorov time $\tau_{\eta}=\sqrt{\nu/\langle\epsilon\rangle}$ and the 
{\em lateral} Lagrangian integral time scale $T_L$ are given. Both are separated by less than an order of magnitude for the
Reynolds number accessible here. Furthermore, the $t^3$-regime extends beyond $T_L$. Figure 2b shows the 
probability density function (PDF) of the {\em lateral} distance at five different times. The PDFs collapse to a 
Richardson-like shape and clearly deviate from the Gaussian limit for times around $T_L$. They cross over to Gaussian shape 
for larger times. A Gaussian distribution, as predicted by Batchelor for the Lagrangian inertial range, is not supported 
by the present study \cite{Batchelor1950}. 

The inset of Fig. 2b unravels another specifics in convection. 
The Lagrangian velocity autocorrelation functions of the vertical and lateral components, $C_{ver}(\tau)$ and $C_{lat}(\tau)$, 
which are calculated along the tracer tracks are shown. The strong anticorrelation of the vertical part 
is due to the constraints in vertical motion, e.g. when the tracers (and the plumes) hit the top and bottom planes.
The pronounced minimum in the vertical correlation is close to a characteristic vertical travel time $\sim H/u_{z,rms}$, where 
$u_{z,rms}=\langle u_z^2\rangle^{1/2}$. The integration of $C_{ver}(\tau)$ with respect to time results in a 
vertical Lagrangian integral time which is $T_L^{(v)}\approx T_L/6$. While the single particle motion decorrelates on a 
shorter time scale as in isotropic turbulence, the lateral pair motion seems to be stronger correlated. The latter is due to 
tracer pair motion in plumes. This circumstence rationalizes the smaller Richardson constant $C_2$ 
and why $R^2(t) \sim t^3$ seems to continue for times $t> T_L$. Our studies demonstrate, that Richardson
dispersion can also be found in anisotropic turbulence. They could thus shed light on why Richardson-like dispersion is frequently 
observed, but the constant $C_2$ strongly scatters around the analytically 
predicted value. 
             
{\em Lagrangian plume detection.} The results on the particle pair dispersion suggest to take a closer look to the 
vertical accelerations and how their amplitudes are correlated with the thermal plumes. In Fig. 3, we show first 
the PDFs of all three acceleration components. As expected, the PDFs of both lateral components, $a_x$ and $a_y$, 
collapse onto each other. Surprisingly, they yield fatter tails compared to the remaining vertical acceleration 
component. The characteristic amplitude for the latter can be estimated by $|a_z|\sim u_{z,rms}/T_L^{(v)}$ which is 
indicated in Fig. 3. The largest vertical acceleration amplitudes result then to about 10 times this estimate. 
The overall symmetric and stretched exponential shape reveals the strong intermittency of Lagrangian motion as in 
isotropic fluid turbulence \cite{Biferale2004,Grauer2007}. The differences in lateral and vertical accelerations
suggest that the rise of buoyant plumes is less gradual than the lateral vortical motion. 
%-------------------------------------------------------------------------------
\begin{figure}
\includegraphics[width=8.5cm]{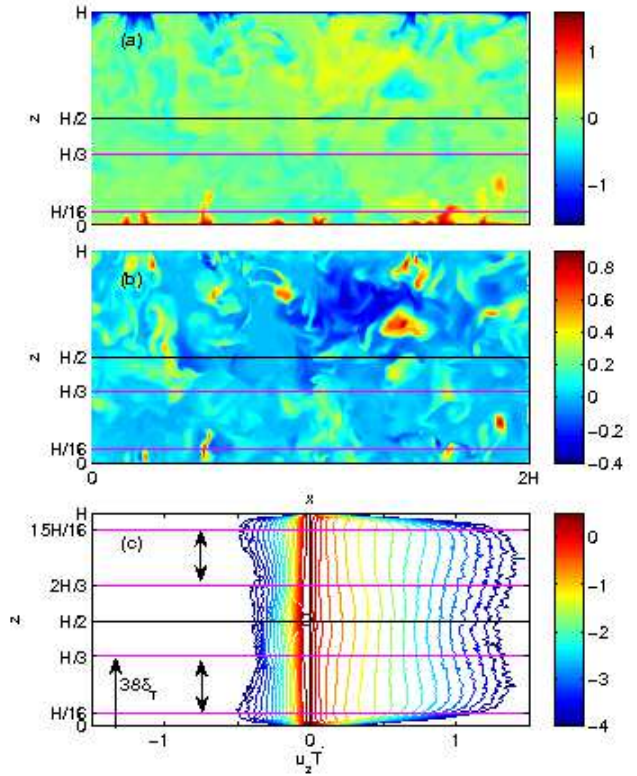}
\caption{(Color online) Lagrangian heat flux as a function of height. (a) Vertical slice cut through the 
instantaneous temperature field $T$. (b) Corresponding product field $u_z T^{\prime}$. (c) Lagrangian probability 
density function (PDF) of $u_z T^{\prime}$ as a function of vertical tracer position $z$. 
The height $z=H/3$ corresponds with 38 times the thermal boundary layer thickness $\delta_T$. The regions between
both thick horizontal lines at $H/3$ and $H/16$ as well as $2H/3$ and $15H/16$ are considered as the mixing zones. 
Color legend is here in decadic logarithm.} 
\label{fig4}
\end{figure}
%--------------------------------------------------------------------------------

It is known that rising or falling plumes can be identified by a positive product of vertical velocity 
and temperature fluctuation, $u_z T^{\prime}>0$ where $T^{\prime}({\bm x},t)=T({\bm x},t)-\langle T\rangle_A(z)$. 
Here, $\langle\cdot\rangle_A$ denotes ensemble averages at fixed height $z$. The product $u_z T^{\prime}$ is 
directly related to the (Lagrangian) Nusselt number \cite{Gasteuil2007} which is given by $Nu_L=1+H/(\kappa\Delta T)
u_{z} T^{\prime}$. Figure 4c shows how probable the positive and negative amplitudes of $u_z T^{\prime}$ as a function 
of height $z$ are. We see, that the distribution is always asymmetric with respect to the axis $u_z T^{\prime}=0$.
This reflects the net transfer of heat from the bottom to the top plane.
The broad support of the PDF is due to the fluctuations of the local heat transfer. The joint PDF shows two features.
Firstly, the support of the PDF for $u_z T^{\prime}>0$ is broadest at about 18 times the thermal boundary layer thickness 
$\delta_T$ (i.e. $z\approx H/6$). Secondly, the support of the PDF for $u_z T^{\prime}<0$ peaks at $H/16$ and decreases
to about $H/3$. In order to shed more light on the structural features which are associated with this statisitics, we show 
in Fig. 4 the temperature $T$ (a) and the product $u_z T^{\prime}$ (b). Clearly, thermal plumes are found were
$u_z T^{\prime}>0$. However, one observes also larger $u_z T^{\prime}<0$ in the vicinity of plumes. Due to the 
incompressibility of the flow, strong upwellings are accompanied by neighboring downwellings which reflects in the 
statistical distribution in the bottom panel of Fig. 4. By combining both aspects, we conclude that the heat transport and
mixing are dominated by thermal plumes in the zone between $H/16\lesssim z\lesssim H/3$. Such plume mixing region has been suggested by Castaing {\it et al.} \cite{Castaing1989} and was determined in an Eulerian analysis \cite{Xia2002}. Compared
to \cite{Xia2002}, the present mixing zone is smaller in extension, but again significantly larger as the thermal boundary layer thickness. This explains why the smart sensor experiment by Gasteuil {\it et al.} \cite{Gasteuil2007} could detect 
successfully the large fluctuations of the heat transfer despite the large probe diameter. 
    
%--------------------------------------------------------------------------------
%\begin{figure}
%\includegraphics[width=6cm]{heatjoint1}
%\caption{(Color online) Contours of the joint Lagrangian PDF of the vertical velocity $u_z$ and the temperature fluctuations 
%$\theta$. Lagrangian properties are probed in three disjoint zones of the convection cell as indicated in each of the panels. 
%The contours of the decadic logarithm of the corresponding joint PDF increment in steps of 0.5 from -5 to 0.5. The first quadrant 
%in the mid panel corresponds with enhanced presence of plumes.} 
%\label{fig5}
%\end{figure}
%--------------------------------------------------------------------------------
\begin{figure}
\includegraphics[width=8cm]{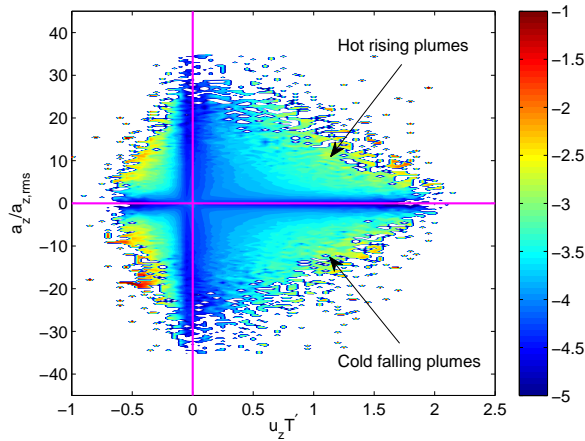}
\caption{(Color online) Joint Lagrangian PDF of vertical acceleration component $a_z/a_{z,rms}$ 
and the product $u_z T^{\prime}$. Color coding is again in decadic logarithmic units as indicated 
by the color bar. The regions for rising hot plumes and falling cold plumes are indicated by arrows.} 
\label{fig6}
\end{figure}
%-------------------------------------------------------------------------------- 
%In Fig. \ref{fig5}, we show the contours of the joint Lagrangian PDF of temperature fluctuation $T^{\prime}$ and 
%vertical velocity $u_z$ for the three zones - boundary zone, mixing zone, and bulk zone as indicated in Fig. 4.
%First, it is seen that the PDF gets increasingly tilted with increasing height which implies the joint 
%appearance of larger vertical velocity amplitudes and smaller temperature fluctuations. In the mixing zone,
%a pronounced deformation to large $T^{\prime}$ and $u_z$ in the highlighted first quadrant is found, a clear
%fingerprint of intensified plume action. 
Finally, the combination of the results on acceleration and local heat transfer result in the joint statistics which is shown
in Fig. \ref{fig6}. In order to highlight the statistical dependence of both fields, we normalize the joint PDF by the single
quantity PDFs, $\Pi(a_z, u_z T^{\prime})=P(a_z, u_z T^{\prime})/[P(a_z) P(u_z T^{\prime})]$.  
The pronounced asymmetry with respect to the axis $u_z T^{\prime}=0$ reflects again the net transfer of heat, similar to Fig.4. 
It can be seen that larger accelerations are associated with rising and falling plumes as well as with resulting down- and
upwellings, respectively. The latter are found in the corresponding diagonal quadrant of the plane. Maxima of $\Pi(a_z, u_z T^{\prime})$ are detected in the outer regions of the support, but not for the strongest acceleration. The figure illustrates thus also which 
fraction of the tail of the acceleration PDF is most probably associated with the plume motion. 

{\em Summary.} Due to vertical
buoyancy driving of convective turbulence, Lagrangian tracer motion is found to be strongly anisotropic in contrast to classical fluid 
turbulence \cite{Mann2000,Boffetta2002}. On the one hand, the vertical Lagrangian transport is found significantly 
shorter-in-time correlated than the lateral one. On the other hand, rising thermal plumes seem to enforce a more coherent lateral motion which results in a smaller Richardson constant for the {\em lateral} Richardson pair dispersion which has been detected here
in an anisotropic flow. Vertical accelerations are diminished compared to lateral ones and their largest amplitudes are not connected with rising and falling plumes. Based on the Lagrangian monitoring of heat transfer, the plume mixing zone has been identified complementary to previous Eulerian analysis. It is believed, that the presented features can be carried over qualitatively to the case of no-slip boundaries. This particular investigation and trends with increasing Rayleigh number are part of our future work.
   
{\em Acknowledgements.}
The author acknowledges support by the Deutsche Forschungsgemeinschaft (DFG) under
grant SCHU 1410/2. The direct numerical simulations have been
carried out at the IBM-p690 cluster JUMP and the IBM Blue Gene/L JUBL at the J\"ulich
Supercomputing Centre (Germany) under grant HIL02.

\end{document}